\documentclass[twocolumn]{emulateapj}

\begin{document}

\title{A \lowercase{$k$}-$\omega$ model for turbulently thermal convection in stars}

\author{Yan Li}
\affil{National Astronomical Observatories/Yunnan Observatory, 
       Chinese Academy of Sciences, P.O. Box 110, Kunming 650011, China\\
       Laboratory for Structure and Evolution of Celestial Objects,
       Chinese Academy of Sciences}
\email{Electronic address: ly@ynao.ac.cn}

\begin{abstract}
Both observations and numerical simulations show that stellar convective 
motions are composed of semi-regular flows of convective rolling cells and the 
fully developed turbulence. Although the convective rolling cells are crucial 
for the properties of the stellar convection that transports heat and mixes 
materials in the stellar interior, their contributions have not been included 
in turbulent convection models proposed up to now. We simplify the structure of 
the convective rolling cells as a cellular pattern moving circle by circle with 
different angular velocities around the center, estimating their typical 
size by solving approximately for the temperature difference over the 
stationary temperature background and their average shear of velocity by 
evaluating approximately their kinetic energy transformed by themselves working 
as thermal engines from the heat involved in the convective rolling cells. We 
obtain a steady state solution in the fully local equilibrium which is similar 
to what is obtained in the standard mixing-length theory, by applying such 
model assumptions to the standard $k$-$\varepsilon$ model and properly choosing 
the model parameter $c_{\varepsilon 3}$. Accordingly, we propose a $k$-$\omega$ 
model to include the transport effect of turbulence in stars. Preliminary 
results of their applications to the sun and other stars with different masses 
and in different evolutionary stages show good agreements with results of the 
standard mixing-length theory and results of numerical simulations for the 
stellar convection.
\end{abstract}


\section{Introduction}

Thermal convection is a common phenomenon in the stellar interior. Convective 
flows are driven in stars by the thermal buoyancy, which results from unstable 
density stratifications. Both observations and numerical simulations show that 
convective motions are characterized by deterministic structures of different 
scales such as unsteady convective rolls and semi-regular convective cells 
created by large-scale instabilities of the buoyancy, along with the fully 
developed turbulence due to extremely high Reynolds numbers in the stellar 
convection zones. The bulk of a convection zone is stochastically filled up 
with numerous unsteady convective rolls pushing each other and leaving little 
space in between \citep{ma07,kh09,tr10}. Close to the boundaries of the 
convective region, semi-regular convective cells jostle onto the surface and 
roughly line up seen for examples as the Rayleigh-B\'{e}nard convection 
\citep{kh00, kh06} and the solar granulation \citep{sn98, fr12}. 

Such semi-deterministic and well-organized structures have already been 
recognized to have fundamental significance on the overall properties of the 
convection. From the dynamical point of view, the buoyancy drives the fluids 
continuously moving up and down alternatively in the convective rolling cells 
by doing the mechanical works respectively on the corresponding flows. Due to 
the equation of mass conservation, the flows form a cellular pattern with the 
warm fluids moving up in the center and the cold fluids moving down along the 
borders. Such a cellular motion creates shears of the velocities between the 
neighbouring layers in the convective rolling cells, which contributes to the 
generation of turbulence. From the thermodynamic point of view, the cellular 
motion introduces inhomogeneous temperature perturbations over the stationary 
temperature background in the convective rolling cells, which results in extra 
fluxes transferring heat into or out of the boundaries of the convective cells. 
Therefore, each of the convective rolling cells runs as an individual heat 
engine, within which fluids absorbing heat at the higher temperature boundary 
while releasing heat at the lower temperature boundary, and the convective 
cell operates cycle by cycle to transform heat into the mechanical work, which 
is the energy supply of the cellular motions themselves in the convective 
rolling cells. 

Thermal convection is an essential input physics for the stellar modelling. It 
not only carries heat outward from the stellar interior, but mixes different 
materials completely in the convective regions as well. In order to incorporate 
these effects of the convection in stellar models, various convection models 
have been proposed, including the most widely used mixing-length theory 
\citep{bv53, bv58} and more recently turbulent convection models 
\citep{xio80, xio89, dxc06, can97, cd98, can01, ly01}. Turbulent convection 
models are based on the averaged hydrodynamic moment equations, and include many 
properties of turbulence such as the local effects of generation, dissipation, 
anisotropy, and the nonlocal effect of turbulent transport. However, all of them 
suffer from a common problem, i.e., the contributions from the convective 
rolling cells have not been considered up to now. Consequently, these models 
either do not result in a steady state solution in the fully local equilibrium, 
or have to adopt a length model of turbulence similar to what has been 
introduced in the standard mixing-length theory. 

In order to incorporate properly the effects of the convective rolling cells 
into the turbulent convection models, we have tried to find out the qualitative 
properties of the convective rolling cells in the stellar convection zone, by 
approximating the configuration of the convective rolling cells based on the 
results of the numerical simulations as a simple eddy structure with fluids 
moving circle by circle around the center of the eddy. The basic equations 
governing the motions of the convective rolling cells are given in Sect. 2. 
In Sect. 3, equations of the standard $k$-$\varepsilon$ model for turbulence 
are introduced firstly, and then their steady state solutions in the fully 
local equilibrium are discussed. Based on the advanced turbulence models, we 
introduce new models for the rate of buoyancy production and for the turbulent 
heat flux with modifications suitable for the turbulent motions in the stellar 
convective regions. With the aid of above models, the temperature gradient in 
the convective region can be derived. In Sect. 4, we first estimate the typical 
size of the convective rolling cells by solving approximately for the 
temperature difference in a convective eddy, and then evaluate the shear 
production rate of turbulence by a model of the velocity shear. For vertical 
flows similar to those down-drafts seen in the numerical simulations of the 
stellar convection, we evaluate their development by introducing a decaying 
distance of the temperature difference along the vertical flow. In Sect. 5, we 
introduce a new choice of parameter $c_{\varepsilon 3}$ in the standard 
$k$-$\varepsilon$ model, which leads to the well known properties of turbulence 
in the stellar convection zones. By comparing with the length model of 
turbulence in the standard mixing-length theory, we introduce a new 
macro-length model for the turbulent convection models. Based on above model 
assumptions, we obtain a steady state solution of the $k$-$\varepsilon$ model 
in the fully local equilibrium, and compare it with the result of the 
mixing-length theory. In accordance with all model assumptions introduced 
above, we suggest a $k$-$\omega$ model for the stellar turbulent convection in 
Sect. 6, and apply it to the sun and some other stellar models with different 
masses in Sect. 7. We summarize in Sect. 8 our main conclusions, and discuss 
some important effects that do not yet properly incorporated in our stellar 
turbulent convection model.

\section{Equations for motions of convective rolling cells}

We approximate the stellar convection zone as a plane parallel structure, and
establish Cartesian coordinates $ (x,y,z) $ with the vertical direction $z$
along the stellar radius. 

We suppose that the stellar turbulent convection is composed of a more or 
less regular motion of convective rolling cells and a fully developed turbulent 
motion. The semi-regular motions of the convective rolling cells are determined 
by the Reynolds-averaged Navier-Stokes equation:
\begin{equation}
\frac{\partial {{U}_{i}}}{\partial t}+{{U}_{j}}
\frac{\partial {{U}_{i}}}{\partial {{x}_{j}}}
={{g}_{i}}-\frac{1}{\rho }\frac{\partial p}{\partial 
{{x}_{i}}}+\frac{\partial }{\partial {{x}_{j}}}\left( \nu \frac{\partial 
{{U}_{i}}}{\partial {{x}_{j}}}-\overline{{{u}_{i}}{{u}_{j}}} \right).
\end{equation}
In Eq. (1), the velocity of the convective flow $ U_i $ and the velocity 
fluctuation of turbulence $ u_i $ can be expressed respectively as: 
\begin{equation}
\begin{array}{cccccc}
 {{U}_{i}} &=& \left( U, V, W \right) , \\
 {{u}_{i}} &=& \left( u, v, w \right) ,
\end{array}
\end{equation}
$ g_i $ the component of gravitational acceleration in the $ x_i $ direction, 
$p$ the pressure, $\rho$ the density, and $\nu$ the molecular viscosity. Note 
that summation should be made over all three components if the subscript $i$ 
appears twice in the same expression. 

The stellar convection zone is as a whole in the static state, satisfying the 
equation of hydrostatic equilibrium:
\begin{equation}
\frac{\partial p}{\partial z}=-{{\rho }_{0}}g,
\end{equation}
where $\rho_0$ is the stationary density distribution in the convection zone 
and the gravitational acceleration $g$ is defined as:
\begin{equation}
{{g}_{i}}=\left( 0,\,0,\,-g \right).
\end{equation}

The unstable stratification results in a density difference $\Delta\rho$ in
the fluid, which is related to the temperature difference $\Delta T$ under
the Boussinesq approximation:
\begin{equation}
\frac{\Delta \rho }{{{\rho }_{0}}}\approx -\beta \Delta T,
\end{equation}
where $T$ is the temperature and the thermodynamic coefficient $ \beta $ is 
defined by:
\begin{equation}
\beta =-\frac{1}{\rho}{{\left( \frac{\partial\rho}{\partial T}\right)}_{p}}.
\end{equation}
Substituting Eqs. (3) and (5) into Eq. (1), the convective motion in the steady
state is described by:
\begin{equation}
U_{j}\frac{\partial U_i}{\partial x_j}
=-\beta g_i \Delta T-\frac{\partial\overline{u_{i} u_{j}}}{\partial x_j}.
\end{equation}
It can be noticed that the first term on the right hand side of Eq. (7) is 
the buoyancy, and the second term describes the resistance due to the Reynolds  
stress.  

The Reynolds-averaged equation of energy conservation for the convective 
rolling cells can be expressed as:
\begin{equation}
\rho T\left( \frac{\partial s}{\partial t}+{{U}_{i}}\frac{\partial s}
{\partial {{x}_{i}}} \right)=\frac{\partial }{\partial {{x}_{i}}}
\left( \lambda \frac{\partial T}{\partial {{x}_{i}}}-F_{C}^{i} \right),
\end{equation}
where $s$ is the entropy of the stellar matter and $ F^i_C $ the convective 
heat flux in the $x_i$ direction. The thermal conductivity of radiation 
$\lambda$ is defined by:
\begin{equation}
\lambda = \frac{ 16 \sigma T^3}{3 \rho \kappa },
\end{equation}
where $\kappa$ is the opacity of the stellar matter and $\sigma$ the 
Stefan-Boltzmann constant. 

Usually the total energy flux $F$ is assumed to be a constant in the stellar 
convective envelope, which results in:
\begin{equation}
F={{F}_{C}}-\lambda \frac{\partial {{T}_{0}}}{\partial z},
\end{equation}
where $T_0$ is the stationary temperature distribution in the convection zone 
and $F_C$ the convective heat flux in the $z$ direction. Substituting Eq. (10) 
into Eq. (8), it can be obtained that the temperature difference is 
approximately determined by:
\begin{equation}
\frac{\partial }{\partial {{x}_{i}}}\left( \lambda \frac{\partial \Delta 
T}{\partial {{x}_{i}}} \right)={{\rho }_{0}}{{c}_{p}}\frac{\partial \Delta 
T}{\partial {{x}_{i}}}{{U}_{i}}+{{\rho }_{0}}{{T}_{0}}\frac{\partial s}
{\partial z}W,
\end{equation}
where $c_p$ is the specific heat at constant pressure. We omit hereafter the 
subscript "0" for the stationary density and temperature, which will give rise 
to no confusion on their meanings.

\begin{deluxetable}{ccccc}
\tabletypesize{\scriptsize}
\tablecaption{Parameters' values of the $ k-\varepsilon $ model}
\tablewidth{0pt}
\tablehead{ \colhead{$c_{\mu}$}          & \colhead{$\sigma_{\varepsilon}$} & 
            \colhead{$c_{\varepsilon1}$} & \colhead{$c_{\varepsilon2}$}     & 
            \colhead{References} }
\startdata
0.09 & 1.3 & 1.44 & 1.92 & \citet{pop} \\
     &     & 1.5  & 2.0  & \citet{ten}  
\enddata
\end{deluxetable}

\section{The \lowercase{$k$}-$\varepsilon$ model for the stellar 
         turbulent convection}

\subsection{Equations of the \lowercase{$k$}-$\varepsilon$ model with buoyancy
            modifications}

The standard $k$-$\varepsilon$ model with buoyancy modifications consists of 
two equations \citep{pop,hos}:
\begin{equation}
\frac{Dk}{Dt}-\frac{\partial}{\partial x_i}
\left( \nu_{t}\frac{\partial k}{\partial x_i} \right) 
=P+G-\varepsilon ,
\end{equation}
\begin{equation}
\frac{D\varepsilon }{Dt}-\frac{\partial }{\partial {{x}_{i}}}
\left( \frac{{{\nu }_{t}}}{{{\sigma }_{\varepsilon }}}\frac{\partial 
\varepsilon }{\partial {{x}_{i}}} \right)={{c}_{\varepsilon 
1}}\left( P+{{c}_{\varepsilon 3}}G \right)\frac{\varepsilon }{k}
-{{c}_{\varepsilon 2}}\frac{{{\varepsilon }^{2}}}{k},
\end{equation}
where $ k=\frac{1}{2}\overline{u_{i}u_{i}} $ is the kinetic energy of 
turbulence, $\varepsilon$ the dissipation rate of $ k $, and $ D/Dt $ the 
co-moving derivative. On the left hand side of Eqs. (12) and (13), the standard 
gradient diffusion hypothesis is adopted to treat the turbulent transport 
process, and the turbulent viscosity $\nu_{t}$ is approximated according to 
the eddy viscosity model as:
\begin{equation}
\nu_{t}=c_{\mu}\frac{k^{2}}{\varepsilon}.
\end{equation}
On the right hand side of Eqs. (12) and (13), $ P $ represents the shear 
production rate of turbulent kinetic energy:
\begin{equation}
P=-\overline{u_{i}u_{j}}\frac{\partial U_{i}}{\partial x_{j}},
\end{equation}
while $G$ describes the contribution from the buoyancy:
\begin{equation}
G=-\beta g_{i}\overline{u_{i}\vartheta},
\end{equation}
where $ \vartheta $ is the temperature fluctuation of turbulence. There are 
some model parameters in above equations, e.g. $ c_{\mu} $, 
$ \sigma_{\varepsilon} $, $ c_{\varepsilon 1} $, $ c_{\varepsilon 2} $,
and $ c_{\varepsilon 3} $. Table 1 lists some choices of their values that
are specified in applications of the standard $ k-\varepsilon $ model.
There are lots of controversy on the value of $ c_{\varepsilon 3} $, and 
we shall discuss this in the following sections. 

\subsection{Steady state solutions in the fully local equilibrium}

If the left hand sides of Eqs. (12) and (13) are equal to zero, the turbulence
is in fully local equilibrium state. This results in a steady state solution: 
\begin{equation}
\frac{P}{\varepsilon}=
 \frac{c_{\varepsilon 2}-c_{\varepsilon 1}c_{\varepsilon 3}}
      {c_{\varepsilon 1}-c_{\varepsilon 1}c_{\varepsilon 3}},
\end{equation}
\begin{equation}
\frac{G}{\varepsilon}=
-\frac{c_{\varepsilon 2}-c_{\varepsilon 1}}
      {c_{\varepsilon 1}-c_{\varepsilon 1}c_{\varepsilon 3}}.
\end{equation}

We use for simplicity the turbulence parameters suggested by \citet{ten} in 
Table 1, and define a new parameter as:
\begin{equation}
c_{\varepsilon 3}
=1+\frac{c_{\varepsilon 2}-c_{\varepsilon 1}}
{c_{\varepsilon 1}}c'_{\varepsilon 3}
=1+\frac{1}{3}c'_{\varepsilon 3}.
\end{equation}
As a result, it can be obtained that:
\begin{equation}
\frac{P}{\varepsilon}=1-\frac{1}{c'_{\varepsilon 3}},
\end{equation}
\begin{equation}
\frac{G}{\varepsilon }=\frac{1}{{{{{c}'}}_{\varepsilon 3}}}.
\end{equation}
It is interesting to notice that only when $c'_{\varepsilon 3}>0$ can the fully
local equilibrium of turbulence appear in the convectively unstable region. 
Furthermore, the condition that $c'_{\varepsilon 3}>1$ must be satisfied to 
ensure a positive rate of shear production.

\subsection{Model for the rate of buoyancy production }

As already pointed out by \citet{kh00} that, in order to reproduce the 
ensemble roll pattern of convective cells, it is crucial to use an algebraic 
model for $ \overline{u_{i}\vartheta} $ including all its production terms: 
\begin{equation}
\left( 1+\frac{1}{y} \right)\overline{{{u}_{i}}\vartheta }=
-{{c}_{\theta }}\frac{k}{\varepsilon }
\left( \frac{T}{{{c}_{p}}}\frac{\partial s}{\partial {{x}_{j}}}
\overline{{{u}_{i}}{{u}_{j}}}+\xi \frac{\partial {{U}_{i}}}{\partial 
{{x}_{j}}}\overline{{{u}_{j}}\vartheta }+\eta \beta 
{{g}_{i}}\overline{{{\vartheta }^{2}}} \right),
\end{equation}
where the turbulent P\'{e}clet number $y$ is defined as:
\begin{equation}
y=\frac{\rho {{c}_{p}}}{\lambda }\frac{{{k}^{2}}}{\varepsilon }.
\end{equation}
On the right hand side of Eq. (22), the first term in the brackets is the 
contribution from the stratification, the second one from the velocity shear 
of convective rolling cells, and the last one from the buoyancy. We introduce 
a modification of time scale from radiative dissipation on the left hand 
side \citep{cd98,ly01}. 

For the auto-correlation of temperature fluctuation $\overline{\vartheta^2}$ 
in Eq. (22), we adopt a local equilibrium approximation that assumes a complete
balance bewteen its production and dissipation:
\begin{equation}
\overline{{{\vartheta }^{2}}}=-{{c'}_{\theta }}\frac{k}{\varepsilon }
\frac{T}{{{c}_{p}}}\frac{\partial s}{\partial {{x}_{i}}}
\overline{{{u}_{i}}\vartheta }.
\end{equation}

Solving for the velocity-temperature correlation $\overline{w\vartheta}$
in the $z$ direction and neglecting some angle-dependent ingredients, we 
finally approximate the rate of buoyancy production by the following model:
\begin{equation}
G=-\frac{{{{{c}'}}_{\mu }}}{1+y^{-1}+{{{{c}'}}_{\mu }}{{c}_{T}}{{\tau }^{2}}
{{N}^{2}}}\frac{{{k}^{2}}}{\varepsilon }{{N}^{2}},
\end{equation}
where the buoyancy frequency $N$ is defined as:
\begin{equation}
{{N}^{2}}=\beta g\frac{T}{{{c}_{p}}}\frac{\partial s}{\partial z},
\end{equation}
and the typical time scale of turbulence $ \tau $ is defined as:
\begin{equation}
\tau=\frac{k}{\varepsilon}.
\end{equation}
It can be noticed from Eq. (25) that the property of the stratification 
determines the effect of buoyancy contribution: $G>0$ in a convection zone 
where $N^2<0$, while $G<0$ in a stably stratified region where $N^2>0$.

\subsection{ Model of the convective heat flux }

Based on similar arguments as discussed in deriving Eq. (25), we approximate 
the convective heat flux $F_C$ by:
\begin{equation}
{{F}_{C}}=-\frac{\rho {{c}_{p}}}{\beta 
g}\frac{{{{{c}'}}_{\mu }}{{\tau }_{h}}k}{1+y^{-1}
+{{{{c}'}}_{\mu }}{{c}_{T}}{{\tau }^{2}}{{N}^{2}}}{{N}^{2}}
\end{equation}
with a characteristic time scale $\tau_h$ that is often approximated by a 
model proposed by \citet{nak} as:
\begin{equation}
{{\tau }_{h}}=\sqrt{{{\tau }_{\theta }}\tau }.
\end{equation}
In Eq. (29), the kinetic time scale of turbulence $\tau$ is defined by 
Eq. (27), and the thermal time scale of turbulence $\tau_\theta$ is 
approximated by:
\begin{equation}
{{\tau }_{\theta }}=\frac{\rho 
{{c}_{p}}}{\lambda }\frac{{{k}^{2}}}{\varepsilon }\tau .
\end{equation}
As a result, the convective heat flux is then expressed as:
\begin{equation}
{{F}_{C}}=-\frac{\lambda }{\beta g}\frac{{{{c'}}_{\mu }}y^{3/2}}
{1+y^{-1}+{{{{c}'}}_{\mu }}{{c}_{T}}{{\tau }^{2}}
{{N}^{2}}}{{N}^{2}}.
\end{equation}

\subsection{The temperature gradient in the stellar convection zone}

In the stellar convective envelope, the temperature gradient $\nabla$ can be 
obtained by substituting Eq. (31) into Eq. (10) as:
\begin{equation}
\nabla =\frac{d\ln T}{d\ln p}={{\nabla }_{r}}+\frac{{{H}_{p}}}{\beta 
gT}\frac{{c'}_{\mu }y^{3/2}}{1+y^{-1}+{{{{c}'}}_{\mu }}{{c}_{T}}{{\tau }^{2}}
{{N}^{2}}}{{N}^{2}},
\end{equation}
where $\nabla_r$ is the radiative temperature gradient assuming that the heat
flux is solely transferred by radiation, and the local pressure scale height 
is defined as:
\begin{equation}
{{H}_{p}}=-\frac{dr}{d\ln p}=\frac{p}{\rho g}.
\end{equation}

The buoyancy frequency $N$ can be expressed as:
\begin{equation}
{{N}^{2}}=-\frac{\beta gT}{{{H}_{p}}}\left( \nabla -{{\nabla }_{ad}} \right),
\end{equation}
where the adiabatic temperature gradient $\nabla_{ad}$ is defined by:
\begin{equation}
{{\nabla }_{ad}}={{\left( \frac{\partial \ln T}{\partial \ln p} \right)}_{s}}.
\end{equation}
With the definition of the efficiency of convective heat transfer $f$ as:
\begin{equation}
f=\frac{\nabla -{{\nabla }_{ad}}}{{{\nabla }_{r}}-{{\nabla }_{ad}}},
\end{equation}
the buoyancy frequency can be written as:
\begin{equation}
{{N}^{2}}=-\frac{\beta gT}{{{H}_{p}}}
\left( {{\nabla }_{r}}-{{\nabla }_{ad}} \right)f=-Ef,
\end{equation}
where $E$ is a function of the stellar structure.

Directly solving for $f$ from Eq. (32) will result in a singular solution. 
Instead, we introduce a new variable:
\begin{equation}
h=\frac{1-f}{f},
\end{equation}
and Eq. (32) becomes:
\begin{eqnarray}
&&\left( 1+y \right){{h}^{2}}+\left( 1+y-{{{{c}'}}_{\mu }}{{y}^{5/2}}
-{{{{c}'}}_{\mu }}{{c}_{T}}E{{\tau }^{2}}y \right)h
-{{{c}'}_{\mu }}{{y}^{5/2}}=0 \nonumber \\
&&\quad{\rm with}\quad  A_1={{{{c}'}}_{\mu }}{{c}_{T}}E
{{\tau }^{2}}y+{{{{c}'}}_{\mu }}{{y}^{5/2}}-y-1.  
\end{eqnarray}
Equation (39) is a quadratic equation and its solution is:
\begin{equation}
h=\frac{A_1 +\sqrt{{A_1^2}+4\left( 1+y\right){{{{c}'}}_{\mu }}
{{y}^{5/2}}}}{2\left( 1+y \right)}. \end{equation}
It is easy to verify that $h$ is always positive not only in the convection
zone ($E>0$) but also in the overshooting regions ($E<0$).

\section{Shear production rate from convective rolling cells}

\subsection{Simplified structure of convective rolling cells}

The velocity structure in a convective rolling cell can be simplified as a 
circular motion in, for example, $(y,z)$ plane:
\begin{equation}
{{U}_{i}}=\left( 0,\,V,\,W \right).
\end{equation}
By use of a system of polar coordinates with the origin at the center of 
the cell, the velocities along $y$ and $z$ directions can be written as:  
\begin{equation}
\begin{array}{ll}
& V=-{{V}_{\theta }}\sin \theta , \\
& W= {{V}_{\theta }}\cos \theta ,
\end{array}
\end{equation}
where the velocity in the $ \theta $ direction can be expressed as:
\begin{equation}
{{V}_{\theta }}=r\Omega (r),
\end{equation}
and the angular velocity $\Omega$ is regarded as only a function of the radius 
$r$ to the center of the cell. It can be noticed that such a velocity 
structure satisfies the continuity equation of an incompressible fluid:
\begin{equation}
\frac{\partial V}{\partial y}+\frac{\partial W}{\partial z}
=\frac{\partial\Omega}{\partial\theta}=0.
\end{equation}

Based on such a simplified structure of convective rolling cells, the shear 
production rate $P$ can be written as:
\begin{equation}
P=-\frac{\partial V}{\partial y}\overline{v^{2}}
  -\frac{\partial W}{\partial z}\overline{w^{2}} 
  -\left( \frac{\partial V}{\partial z}
         +\frac{\partial W}{\partial y} \right) \overline{vw}.
\end{equation}
If the Reynolds stress is approximated by the simple eddy viscosity model:
\begin{equation}
\overline{u_{i}u_{j}}=\frac{2}{3}k\delta_{ij}
-\nu_{t}\left( \frac{\partial U_{i}}{\partial x_{j}}
              +\frac{\partial U_{j}}{\partial x_{i}} \right),
\end{equation}
we obtain by substituting Eqs. (42) and (46) into Eq. (45) that:
\begin{equation}
P=\nu_{t}S^2,
\end{equation}
where the shear of velocity $S$ in the convective cell is defined by:  
\begin{equation}
S=r\frac{\partial\Omega}{\partial r}.
\end{equation}
It can be noticed that only the velocity shear contributes to the production of
turbulence, while rigid body rotation has no effect on turbulence.

\subsection{The temperature difference in a convective cell}

The motion of a convective rolling cell will result in a temperature
difference $\Delta T$ superimposed on the stationary temperature background.
For a circular motion considered above, $\Delta T$ approximately 
satisfies the following equation in the steady state:
\begin{equation}
\frac{\lambda }{\rho {{c}_{p}}}\left[ \frac{1}{r}
\frac{\partial }{\partial r}\left( r\frac{\partial \Delta T}{\partial r} 
\right)+\frac{1}{{{r}^{2}}}\frac{{{\partial }^{2}}\Delta T}{\partial 
{{\theta }^{2}}} \right]-\frac{{{V}_{\theta }}}{r}\frac{\partial \Delta 
T}{\partial \theta }=\frac{{{N}^{2}}}{\beta g}{{V}_{\theta }}\cos \theta .
\end{equation}
The first term on the left hand side of Eq. (49) describes heat dissipation due
to radiation, and the second term gives the temperature difference of a
displacement over a stratified fluid.

If the heat dissipation of radiation is ignored, the solution of Eq. (49) is:
\begin{equation}
\Delta T=-\frac{{{N}^{2}}}{\beta g}r\sin \theta =-\frac{{{N}^{2}}}{\beta g}z,
\end{equation}
which is exactly what the standard mixing-length theory adopts. The general 
solution of Eq. (49) can thus be written in the complex form as:
\begin{equation}
\Delta T=i\frac{{{N}^{2}}}{\beta g}
\left[ r+B\left( r \right) \right]{{e}^{i\theta }},
\end{equation}
where $B$ is assumed to be a function of the radius $r$. Substituting Eq. (51) 
into Eq. (49) and introducing a new independent variable:
\begin{equation}
\zeta =r\sqrt{i\frac{\rho {{c}_{p}}\Omega }{\lambda }},
\end{equation}
we obtain:
\begin{equation}
\zeta^2\frac{\partial^2 B}{\partial\zeta^2}+\zeta\frac{\partial B}
{\partial\zeta}-\left( 1+\zeta^2 \right) B=0.
\end{equation}
This is a first-order modified Bessel equation, and its solution is:
\begin{equation}
B={{B}_{0}}{{I}_{1}}(\zeta )={{B}_{0}}\sum\limits_{m=0}^{\infty }
{\frac{1}{m!(m+1)!}}{{\left( \frac{\zeta }{2} \right)}^{2m+1}}.
\end{equation}
Therefore, the solution of temperature difference can be approximately 
written as:
\begin{equation}
\Delta T=i\frac{{{N}^{2}}{{R}_{b}}}{\beta g}\left[ \frac{r}{{{R}_{b}}}
-{B_0}{I_1} \left( \frac{r}{{{R}_{b}}} \right) \right]{{e}^{i\theta }}.
\end{equation}
This solution satisfies the boundary condition that $\Delta T=0$ at $r=0$. 
On the other side, $\Delta T$ should be zero at the surface of a convective 
rolling cell, which defines a typical size of convective rolling cells 
$R_b$ as:
\begin{equation}
{{R}_{b}}=\sqrt{\frac{\lambda }{\rho {{c}_{p}}\Omega }}.
\end{equation}

From the thermodynamic point of view, however, fluids undergoing the circular 
motion in a convective rolling cell operate as a heat engine in the convection 
zone, absorbing heat when it moves inward to higher temperature region and 
releasing heat when it moves outward to lower temperature region and converting 
the difference of heat it absorbs and releases into its kinetic energy. As a 
result, the kinetic energy of its circular motion should be proportional to 
the heat involved in it:  
\begin{equation}
V_{\theta }^{2}\approx {{\Omega }^{2}}R_{b}^{2}\approx {{c}_{p}}\Delta T.
\end{equation}
Considering Eqs. (55), (56), and (57) and assuming the turbulence in the fully 
local equilibrium described by Eq. (21), we approximate the averaged angular 
velocity of the circular motion of the convective cells by:
\begin{equation}
{{\Omega }^{3/2}}=\sqrt{\frac{\rho {{c}_{p}}}{\lambda }}\frac{\alpha 
{{H}_{p}}}{{{{{c}'}}_{\mu }}{{{{c}'}}_{\varepsilon 3}}{{\tau }^{2}}},
\end{equation}
where $ \alpha $ is a parameter similar as in the mixing-length theory.

\subsection{The averaged shear of velocity in convective rolling cells}

Usually, the unstable stratification in the stellar convection zone is 
fairly weak, and the circular motion reaches a steady state in a convective 
cell if the Reynolds stress balances the centrifugal force of circular 
motion. With the aid of Eqs. (7) and (42), it can be obtained:
\begin{equation}
\Omega^{2}r\approx 
 \left( \frac{\partial\overline{v^2}}{\partial y}
       +\frac{\partial\overline{vw} }{\partial z}\right)\cos\theta 
+\left( \frac{\partial\overline{w^2}}{\partial z}
       +\frac{\partial\overline{vw} }{\partial y}\right)\sin\theta.
\end{equation}
Substituting the eddy viscosity model Eq. (46) into Eq. (59) and assuming that 
the shear of the circular motion is nearly a constant, we obtain:
\begin{eqnarray}
{{\Omega }^{2}}r &\approx& \frac{2}{3}\frac{\partial k}{\partial r}
  -2\left( \frac{\partial {{\nu }_{t}}}{\partial y}
           \frac{\partial V}{\partial y}\cos \theta 
          +\frac{\partial {{\nu }_{t}}}{\partial z}
           \frac{\partial W}{\partial z}\sin \theta  \right) \nonumber\\
&+&\left(  \frac{\partial V}{\partial z}
          +\frac{\partial W}{\partial y} \right)
\left(     \frac{\partial {{\nu }_{t}}}{\partial z}\cos 
\theta    +\frac{\partial {{\nu }_{t}}}{\partial y}\sin \theta  \right).
\end{eqnarray}
It is interesting to note that an axially symmetric circular motion can
induce an azimuth-dependent turbulence. Averaging Eq. (60) over all azimuthal 
angle $ \theta $ and taking the averaged size of convective cells into 
account, we can evaluate the averaged kinetic energy of turbulence in a 
convective cell as:
\begin{equation}
k\approx\frac{3}{4}\Omega^{2}r^{2}\approx\Omega^{2}R^{2}_b.
\end{equation}
It can be seen that the kinetic energy is approximately partitioned more or 
less equally in the circular motion and in turbulence. 

The averaged shear of velocity in the convective rolling cells can be regarded 
as the ratio of their typical velocity to their typical size. 
It should be recognized that the length scale of thermal convection is 
not only determined by the averaged size of convective cells, but also 
restricted by the macro-length scale of turbulence. When the size of the 
convective cell is larger than the macro-length of turbulence, the flow pattern 
of thermal convection is dominated by a cellular structure fully clogged up 
with numerous convective cells. If the macro-length of turbulence is larger 
than the averaged size of convective cells, the structure of an individual 
convective cell may be destroyed and several convective cells will merge into 
a larger and somewhat chaotic flow pattern with a length scale comparable to 
the macro-length of turbulence. 

Based on above arguments, we assume that the averaged shear of velocity in the 
convective cells can be approximated by:
\begin{equation}
{{S}^{2}}\propto \frac{V_{\theta }^{2}}{R_{b}^{2}+{{L}^{2}}},
\end{equation}
where $ L $ is the macro-length of turbulence. It can be noticed that: 
\begin{equation}
{{\tau }^{2}}{{S}^{2}}\to \left\{ 
\begin{array}{ll}
 {{\tau }^2}{L^{-2}}k\qquad      & {\rm{if}}\quad {{R}_{b}}\ll L , \\
 0                               & {\rm{if}}\quad {{R}_{b}}\gg L .
\end{array} \right.
\end{equation}
Accordingly, the velocity shear dominates the generation of turbulence for
small convective cells, while the buoyancy production plays the major role
to produce turbulence for large convective cells within them the temperature 
difference is able to become large enough. As a result, we suggest a model 
for the averaged shear of velocity in the convective cells as:
\begin{equation}
{{\tau }^{2}}{{S}^{2}}=\frac{1}{c_L+\left( L\Omega \right)^{-2}k}
\end{equation}
with $ c_L=(c_\mu)^{3/4} $. 

\subsection{Decay of the temperature difference along the vertical direction}

As already discussed above, the convective rolling cells can effectively 
restrict the macro-length of turbulence, especially for the case when the
macro-length of turbulence is much longer than the size of a convective cell. 
It is therefore interesting to understand how a plume-like vertical flow 
that is formed by merging several convective cells restricts the development 
of the macro-length of turbulence in the $z$ direction, or, in other words, how 
far a convective rolling cell can move in the vertical direction. 

The buoyancy is determined by the temperature difference of convective flow 
over the stationary structure. We assume that the velocity of a vertical plume 
is characterized by:
\begin{equation}
{{U}_{i}}=\left( 0,\,0,\,W \right),
\end{equation}
where the velocity $W$ in the $z$ direction is only a function of distance $r$ 
to axis $z$ in the horizontal plane $(x, y)$ as:
\begin{equation}
W=W(x,y)=W(r),
\end{equation}
which is similar to the case that axis $z$ is not placed at the center of a 
convective cell but at the boundary of two adjacent convective cells.

The equation of energy conservation Eq. (11) is now written as:
\begin{equation}
\frac{\lambda}{\rho c_p}\left[ \frac{1}{r}\frac{\partial}{\partial r} 
\left( r\frac{\partial\Delta T}{\partial r} \right) 
+\frac{\partial^2 \Delta T}{\partial z^2} \right] 
=\frac{\partial\Delta T}{\partial z}W+\frac{N^2}{\beta g}W.
\end{equation}
The general solution of Eq. (67) is of the form:
\begin{equation}
\Delta T=-\frac{{{N}^{2}}}{\beta g}
\left[ z+Q\left( r \right) H\left( z \right) \right],
\end{equation}
where the first term in the bracket of the right hand side is the well known 
mixing-length solution that has already been discussed for the convective cells 
in Eq. (50), and the second term is what we are interested in at present. 
Substituting the formal solution Eq. (68) into the differential equation 
Eq. (67), we obtain:
\begin{equation}
\frac{\lambda }{\rho {{c}_{p}}}\left[ \frac{H}{r}\frac{\partial }{\partial r}
\left( r\frac{\partial Q}{\partial r}\right)+Q\frac{{{\partial }^{2}}H}
{\partial {{z}^{2}}} \right]=WQ\frac{\partial H}{\partial z}.
\end{equation}

Equation (69) can be solved by the method of separation of variables:
\begin{equation}
\frac{1}{r}\frac{d}{dr}\left( r\frac{dQ}{dr} \right)-\frac{Q}{R_{b}^{2}}=0,
\end{equation}
\begin{equation}
\frac{{{d}^{2}}H}{d{{z}^{2}}}-\frac{1}{{{H}_{b}}}\frac{dH}{dz}
+\frac{H}{R_{b}^{2}}=0,
\end{equation}
where
\begin{equation}
H_b=\frac{\lambda}{\rho c_p W}.
\end{equation}
The solution of Eq. (70) is a zero order modified Bessel function, which 
describes an exponentially decay of velocity from the center of the plume 
and $R_b$ measures therefore the width of the plume or the size of the 
convective cell. 

With the aid of a new variable:
\begin{equation}
z=R_b \zeta,
\end{equation}
Eq. (71) can be written as:
\begin{equation}
\frac{{{d}^{2}}H}{d{{\zeta }^{2}}}-\chi \frac{dH}{d\zeta }+H=0,
\end{equation}
where
\begin{equation}
\chi=\frac{R_b}{H_b}.
\end{equation}
This is a damped oscillation equation, and its solution in the weak damping 
limit ( $\chi\ll 1$ ) is:
\begin{equation}
H=H_0 e^{-z/2H_b} \cos\left( \frac{z}{R_b}+\phi_0 \right).
\end{equation}
It is interesting to note that the temperature difference along the $z$ 
direction is a damped oscillating distribution, with the size of each unit 
being similar to a convective cell and a decaying distance of $H_b$. 

On the other hand, the kinetic energy of a convective plume is simply a 
result of the work done by the buoyancy:
\begin{equation}
W^2\approx-N^2 R^2_b,
\end{equation}
because the buoyancy can only act within a distance comparable to the averaged 
size of convective cells. Considering Eqs. (56) and (72) and assuming that the 
turbulence is in the local equilibrium state, the decaying length of the
temperature difference along a vertical plume can be approximately estimated 
as:
\begin{equation}
H_{b}^{2}\approx {{{c}'}_{\mu }}{{{c}'}_{\varepsilon 3}}
{{\Omega }^{2}}{{\tau }^{2}}R_{b}^{2}.
\end{equation}

\section{Local solutions of the \lowercase{$k$}-$\varepsilon$ model}

\subsection{Choice of model parameter $ c'_{\varepsilon 3} $ }

Usually, the model parameter $ c_{\varepsilon 3} $ is taken to be a constant
in the literature, and choices of its value disperse in a wide range from 
-1.4 to 1.45 \citep{bau}. As already discussed in Sect. 3, the condition 
that $ c'_{\varepsilon 3}>1 $ should be satisfied to ensure a state of
local equilibrium existing in an unstably stratified region. Accordingly, we 
suggest a reasonable choice of $ c'_{\varepsilon 3} $ for the stellar turbulent 
convection as:
\begin{equation}
{{{c}'}_{\varepsilon 3}}=1+\frac{{{c}_{\mu }}}{{{c}_{L}}-{{c}_{\mu }}
+{{\left( \sqrt{{{{{c}'}}_{\mu }}{{{{c}'}}_{\varepsilon 3}}}
{{\tau }^{2}}{{\Omega }^{2}} \right)}^{-1}}}.
\end{equation}
It is therefore easy to see that:
\begin{equation}
1<{{{c}'}_{\varepsilon 3}}<1+\frac{{{c}_{\mu }}}{{{c}_{L}}-{{c}_{\mu }}},
\end{equation}
and with the aid of Eq. (19) that:
\begin{equation}
\frac{4}{3}<{{c}_{\varepsilon 3}}<\frac{4}{3}
+\frac{1}{3}\frac{{{c}_{\mu }}}{{{c}_{L}}-{{c}_{\mu }}}.
\end{equation}

Using the model of averaged velocity shear Eq. (64) and substituting Eq. (79) 
into Eq. (20), we obtain:
\begin{equation}
{{L}^{2}}=\sqrt{{{{{c}'}}_{\mu }}{{{{c}'}}_{\varepsilon 3}}}
\frac{{{k}^{3}}}{{{\varepsilon }^{2}}},
\end{equation}
which confirms the assumption that $L$ in the model of averaged velocity 
shear is indeed the macro-length of turbulence in the state of fully 
local equilibrium. On the other hand, substituting Eq. (79) into the local 
equilibrium condition of buoyancy Eq. (21) and using Eq. (82), we obtain:
\begin{equation}
k\approx -\sqrt{{{{c'}}_{\mu }}{{{c'}}_{\varepsilon 3}}}{{L}^{2}}{{N}^{2}},
\end{equation}
which is similar to the corresponding equation that has been used in the 
standard mixing-length theory. 

\subsection{Macro-length model in the mixing-length theory}

In order to close the turbulent convection model discussed above, we have to 
specify the macro-length of convective turbulence. From Eq. (83) we know that
the production of turbulent kinetic energy is a direct result of the 
mechanical work done by the buoyancy. The buoyancy comes from the temperature 
difference between the turbulently convective flow and its surroundings. As a 
result, the mechanical work done by the buoyancy is closely related to the 
amount of heat enclosed in the turbulent flow:
\begin{equation}
k\approx {c_p}\Delta T\approx -{c_p}\frac{{{N}^{2}}}{\beta g}{{L}_{b}},
\end{equation}
where $ L_b $ is an effective length scale over which the buoyancy 
efficiently does work on the turbulent flow in the stellar convection zone. 
By use of Eqs. (83) and (84) we obtain:
\begin{equation}
{{L}^{2}}\approx {{H}_{p}}{{L}_{b}}.
\end{equation}

If the effects of the convective rolling cells are not taken into account, it 
is reasonable to assume that $ L_b $ is just the macro-length of turbulence 
$ L $ itself. It can be seen from Eq. (85) that:
\begin{equation}
L\approx {{H}_{p}},
\end{equation}
which is exactly what the standard mixing-length theory assumes. 

\subsection{Macro-length model for stellar turbulent convection}

One of the major effects of the convective rolling cells is to restrict the
effective length for the buoyancy doing work on the turbulent flow. Assuming 
that $ L_b $ can be approximated by the decaying length of temperature 
difference along the $z$ direction $ H_b $, we obtain:
\begin{equation}
{{L}^{2}}\approx \sqrt{{{{{c}'}}_{\mu }}{{{{c}'}}_{\varepsilon 3}}} 
\tau\Omega {{R}_{b}}{{H}_{p}}.
\end{equation}
With the aid of Eqs. (56) and (87), we suggest a model for the macro-length of 
turbulence in the stellar convection zone as:
\begin{equation}
L^2=\alpha {H_p}\sqrt{\tau \Omega \frac{\lambda \tau }{\rho {c_p}}},
\end{equation}
where $ \alpha $ is a parameter similar as in the mixing-length theory. This
macro-length model of turbulence will be applied to regions in both unstable 
and stable stratification. 

\subsection{Steady state solutions in the fully local equilibrium}

In the local equilibrium state, it can be found from Eqs. (58), (82), and (88) 
that:
\begin{equation}
{{\tau }^{2}}={{\left( \frac{\rho {{c}_{p}}}{\lambda } \right)}^{2}}
{{\left( \frac{\alpha {{H}_{p}}}{{{{{c}'}}_{\mu }}{{{{c}'}}_{\varepsilon 3}}} 
\right)}^{4}}\frac{{{\left( {{{{c}'}}_{\mu }}{{{{c}'}}_{\varepsilon 3}} 
\right)}^{3/2}}}{{{y}^{3}}}.
\end{equation}
Substituting Eqs. (58) and (89) into Eq. (79), the model parameter 
${c'}_{\varepsilon 3}$ can be rewritten in the local equilibrium state as:
\begin{equation}
{c'}_{\varepsilon 3}=1+\frac{c_{\mu }}{{c_L}-{c_{\mu }}+{{y}^{-1}}}.
\end{equation}
With the definition of a dimensionless quantity $x$ as:
\begin{equation}
\frac{1}{x}={{\left( {{c'}_{\mu }}{{c'}_{\varepsilon 3}} \right)}^{3/4}}
\frac{\rho {{c}_{p}}}{\lambda }{{\left( \frac{\alpha {{H}_{p}}}
{{{{{c}'}}_{\mu }}{{{{c}'}}_{\varepsilon 3}}}\right)}^{2}}\sqrt{\frac{\beta 
gT}{{{H}_{p}}}\left( {{\nabla }_{r}}-{{\nabla }_{ad}} \right)},
\end{equation}
Eq. (40) can be rewritten as:
\begin{eqnarray}
h=\frac{A_2 +\sqrt{{A_2^2}+4\left( 1+y\right){{{{c}'}}_{\mu }}
{{y}^{5/2}}}}{2\left( 1+y \right)} \quad\quad \\
{\rm with}\quad  A_2={{{{c}'}}_{\mu }}{{c}_{T}}(xy)^{-2}
+{{{{c}'}}_{\mu }}{{y}^{5/2}}-y-1. \nonumber
\end{eqnarray}

Combining Eqs. (21) and (37), the steady state solution in the fully local 
equilibrium is given by:
\begin{equation}
1+y=\left( {{c}_{T}}+{{{{c}'}}_{\varepsilon 3}} 
\right)\frac{{{{{c}'}}_{\mu }}f}{{{x}^{2}}{{y}^{2}}}.
\end{equation}
By use of Eq. (32), it can be found that:
\begin{equation}
1+y+\left( 1+\frac{{{c}_{T}}}{{{{{c}'}}_{\varepsilon 3}}} 
\right){{{c}'}_{\mu }}{{y}^{5/2}}=\left( {{c}_{T}}+{{{{c}'}}_{\varepsilon 3}} 
\right)\frac{{{{{c}'}}_{\mu }}}{{{x}^{2}}{{y}^{2}}}.
\end{equation}
Furthermore, it can be found by combining Eqs. (93) and (94) that:
\begin{equation}
h=\left( 1+\frac{{{c}_{T}}}{{{{{c}'}}_{\varepsilon 3}}} 
\right)\frac{{{{{c}'}}_{\mu }}{{y}^{5/2}}}{1+y}.
\end{equation}

\subsection{Comparisons with the standard mixing-length theory}

According to the mixing-length theory, the macro-length of convection is
characterized by the local pressure scale height, e.g. Eq. (86). As a result,
we can approximate the turbulent time scale $\tau$ as:
\begin{equation}
{{\tau }^{2}}\approx {{\left( \frac{\rho {{c}_{p}}}{\lambda } \right)}^{2}}
{{\left( \frac{\alpha {{H}_{p}}}{{{{{c}'}}_{\mu }}{{{{c}'}}_{\varepsilon 3}}} 
\right)}^{4}}\frac{{{\left( {{{{c}'}}_{\mu }}{{{{c}'}}_{\varepsilon 3}} 
\right)}^{3/2}}}{{{y}^{2}}}.
\end{equation}
Substituting Eq. (96) into the local equilibrium condition Eq. (21), we obtain:
\begin{equation}
1+y=\left( {{c}_{T}}+{{{{c}'}}_{\varepsilon 3}} \right)
\frac{{{{{c}'}}_{\mu }}f}{{{x}^{2}}y}.
\end{equation}
The solution of Eq. (97) is:
\begin{equation}
2xy=\sqrt{{{x}^{2}}+4\left( {{c}_{T}}+{{{{c}'}}_{\varepsilon 3}} 
\right){{{{c}'}}_{\mu }}f}-x.
\end{equation}

On the other hand, we assume that the time scale $\tau_h$ of the convective 
heat flux is proportional to the time scale of turbulence $\tau$ in Eq. (29). 
As a result, we obtain that:
\begin{equation}
\left( {{c}_{T}}+{{{{c}'}}_{\varepsilon 3}}\right)\frac{{{{{c}'}}_{\mu }}}
{{{x}^{2}}y}=1+y+\left( 1+\frac{{{c}_{T}}}{{{{{c}'}}_{\varepsilon 3}}} 
\right){{{c}'}_{\mu }}{{y}^{2}}.
\end{equation}
Substituting Eq. (97) into Eq. (99) we obtain:
\begin{equation}
{x^3}{y^3}=\left( {c_T}+{{c'}_{\varepsilon 3}} \right){{\left( 
1+\frac{c_T}{{c'}_{\varepsilon 3}} \right)}^{-1}}x\left( 1-f \right).
\end{equation}
Substituting Eq. (100) into Eq. (98) we finally obtain:
\begin{equation}
{{\left( \sqrt{{{x}^{2}}+4\left( {{c}_{T}}+{{{{c}'}}_{\varepsilon 3}} 
\right){{{{c}'}}_{\mu }}f}-x \right)}^{3}}
=\frac{8\left( {{c}_{T}}+{{{{c}'}}_{\varepsilon 3}} \right)}
{1+{c_T}/{{c'}_{\varepsilon 3}}}x\left( 1-f \right).
\end{equation}
It can be seen that except for different model parameters, Eq. (101) is exactly
the same as the result of the standard mixing-length theory \citep{cox}:
\begin{equation}
{{\left( \sqrt{f+{{x}^{2}}}-x \right)}^{3}}=\frac{8}{9}x\left( 1-f \right).
\end{equation}

Our results are compared with that of the mixing-length theory in Fig. 1, in 
which we use $h$ defined by Eq. (38) as the dependent variable in order to show 
more clearly the asymptotic behaviours of the two convection models. It can 
be seen that our model of Eq. (94) has the same asymptotic behaviour of a -2/3
decreasing index as that of the mixing-length theory when $x \to 0$, while 
declines more slowly with a -5/2 decreasing index than the mixing-length
theory of a -4 decreasing index when $x \to \infty $.

\begin{figure}
\epsscale{1.2}
\plotone{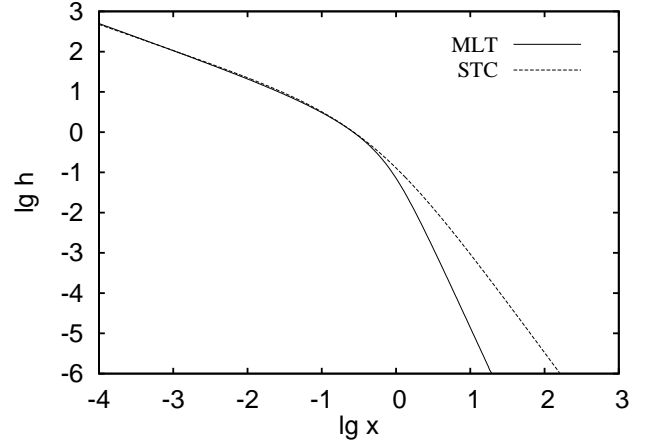}
\caption{The efficiency of the convective heat transfer based on different convection models. The solid line (MLT) represents the result of the standard mixing-length theory, and the dashed line (STC) is the result of the steady state solution of the $k$-$\varepsilon$ model in the fully local equilibrium. }
\end{figure}

\section{A \lowercase{$k$}-$\omega$ model for stellar turbulent convection}

Based on the model assumptions discussed in the preceding sections, the 
equation for the dissipation rate of the kinetic energy of turbulence can be
written as:
\begin{eqnarray}
&& \frac{D\varepsilon }{Dt}-\frac{\partial }{\partial {{x}_{i}}}
\left( \frac{{{\nu }_{t}}}{{{\sigma }_{\varepsilon }}}\frac{\partial 
\varepsilon }{\partial {{x}_{i}}} \right) \nonumber\\
&=&\frac{3}{2}\left[ P+\left( \frac{4}{3}+\frac{1}{3}
\left( {c'}_{\varepsilon 3}-1 \right)  \right)G \right]
\frac{\varepsilon }{k}-2\frac{{{\varepsilon }^{2}}}{k} \nonumber\\ 
&=&\left( 2+\frac{1}{2} \left( {c'}_{\varepsilon 3}-1 \right) \right) 
\left[ \frac{Dk}{Dt}-\frac{\partial }{\partial {{x}_{k}}}
\left( {{c}_{\mu }}\frac{{{k}^{2}}}{\varepsilon}\frac{\partial k} 
{\partial {x_k}} \right) \right]\frac{\varepsilon }{k} \nonumber\\ 
&-&\frac{P}{2{{\Omega }^{2}}\sqrt{{{{{c}'}}_{\mu }}{{{{c}'}}_{\varepsilon 
3}}}}\frac{{{\tau }^{-2}}-\sqrt{{{{{c}'}}_{\mu }}{{{{c}'}}_{\varepsilon 
3}}}{{L}^{-2}}k}{{{c}_{L}}-{{c}_{\mu }}+{{\left( \sqrt{{{{{c}'}}_{\mu }}
{{{{c}'}}_{\varepsilon 3}}}{{\tau }^{2}}{{\Omega }^{2}} 
\right)}^{-1}}}\frac{\varepsilon }{k}. 
\end{eqnarray}

There are two facts to be noticed in Eq. (103). The first term on the right 
hand side of Eq. (103) contributes to an additional diffusion of turbulence as
already pointed out by \citet{pop}. The second term provides a local 
equilibrium state of Eq. (103) for not only the unstable stratification in the 
stellar convection zone but also the stable stratification in the overshooting 
regions. In consistency to the quasi-equilibrium assumption used to derive
the macro-length model of turbulence, we introduce a model equation for the
turbulence frequency $\omega$ as:
\begin{eqnarray}
\frac{D\omega }{Dt}&-&\frac{\partial }{\partial {{x}_{i}}}\left( 
\frac{{{\nu }_{t}}}{{{\sigma }_{\omega }}}\frac{\partial \omega }
{\partial {x_i}} \right)=\sqrt{{{c'}_{\mu }}{{c'}_{\varepsilon 3}}}
\frac{k}{L^2}-{{\omega }^2} \nonumber\\
&=&\frac{{D^{1/3}}{H^{-4/3}}}{\sqrt{{{c'}_{\mu }}
{{c'}_{\varepsilon 3}}}}k{{\omega }^{1/3}}-{{\omega }^{2}},
\end{eqnarray}
where the turbulence frequency $\omega$ is defined by:
\begin{equation}
\omega =\frac{1}{\tau }=\frac{\varepsilon }{k},
\end{equation}
and two quantities of the stellar structure are defined by:
\begin{equation}
D=\frac{\rho {{c}_{p}}}{\lambda }
\end{equation}
and
\begin{equation}
H=\frac{\alpha {H_p}}{{{c'}_{\mu }}{{c'}_{\varepsilon 3}}}.
\end{equation}
We adopt Eq. (90) for the model parameter ${c'}_{\varepsilon 3}$ in accordance
with the quasi-equilibrium assumption used to derive Eq. (104).

By use of Eqs. (37) and (40), the production rate of buoyancy can be written 
as:
\begin{eqnarray}
G=\frac{2{{c'}_{\mu }}Ey}{A_3+\sqrt{{A_3^2}
+4{{c'}_{\mu }}{{y}^{5/2}}{{c'}_{\mu }}
{c_T}E{{\tau }^{2}}y}}\frac{{{k}^{2}}}{\varepsilon } \quad\\
{\rm with}\quad A_3=1+y+{{c'}_{\mu }}{{y}^{5/2}}-{{c'}_{\mu }}
{c_T}E{{\tau }^{2}}y \quad \nonumber.
\end{eqnarray}
As a result, the kinetic energy equation of turbulence can be written as:
\begin{eqnarray}
&& \frac{Dk}{Dt}-\frac{\partial }{\partial {{x}_{i}}}\left( 
{{\nu }_{t}}\frac{\partial k}{\partial {{x}_{i}}} \right) \nonumber\\
&=&\frac{{{c'}_{\mu }}ED{{k}^{2}}\omega }{{{\omega }^{3}}+Dk{{\omega }^{2}}
+{{c'}_{\mu }}{{\left( Dk\right)}^{5/2}}{{\omega }^{1/2}}-{{c'}_{\mu }}
{{c}_{T}}EDk}\frac{2}{1+\sqrt{1+A}}  \nonumber\\
&-&\frac{\left( {c_L}-{c_{\mu }} \right){{c'}_{\mu }}{c'}_{\varepsilon 3}
{H^{8/3}}{D^{1/3}}{{\omega }^{7/3}}+k}{{c_L}{c'}_{\mu }{c'}_{\varepsilon 3}
{{H}^{8/3}}{{D}^{1/3}}{{\omega }^{7/3}}+k}k\omega, \\
& &{\rm where}\nonumber\\
&A&=\frac{4{{c'}_{\mu }}{{\left( Dk \right)}^{5/2}}{{\omega }^{1/2}}
{{c'}_{\mu }}{{c}_{T}}EDk}{{{\left[ {{\omega }^{3}}+Dk{{\omega }^{2}}
+{{c'}_{\mu }}{{\left( Dk \right)}^{5/2}}{{\omega }^{1/2}}
-{{c'}_{\mu }}{c_T}EDk \right]}^{2}}}\nonumber.
\end{eqnarray}

\section{Applications of the \lowercase{$k$}-$\omega$ model in different stellar 
         models}

In order to compare the stellar turbulent convection model proposed in this
paper with the standard mixing-length theory, we have applied both of them in
calculations of stellar evolutionary models and compared the results. Our 
stellar evolutionary models are computed by a stellar evolution code h04.f 
originally described by Paczynski and Kozlowski and updated by \citet{sie}. 
The OPAL opacities \citep{rai,iar} are used in the high temperature region 
and the opacities from molecules and grains \citep{alx} are used in the low 
temperature region. The Livermore Laboratory equation of state \citep {rsi} 
are adopted in our calculations. The nuclear reaction rates are updated 
according to \citet{bap} and \citet{har}. We have modified the treatment of 
element diffusion according to \citet{tbl}. 

\subsection{Comparisons of solar models}

Firstly, we apply the steady state solution (e.g. Eq. (94) ) of the 
$k$-$\omega$ model in the fully local equilibrium to construct solar models. In 
order to calibrate our solar models to achieve the observed solar parameters 
(solar age $\tau_{\odot}=4.57\,$Gyr, solar mass 
$M_{\odot}=1.9891\times 10^{33}\,$g, solar luminosity 
$L_{\odot}=3.839\times 10^{33}\,{\rm erg\,s^{-1}}$ \citep{bap},
and solar radius $R_{\odot}=6.9566\times 10^{10}\,$cm \citep{hab}),
we adjust initial helium abundance $Y_0$ and the mixing-length parameter 
$\alpha$ iteratively to ensure a relative accuracy of $10^{-4}$ for those 
solar parameters. The values of other parameters of the turbulent convection 
model are summarized in Table 2, as well as some basic properties of our 
solar models. In order for our solar models to have a depth of the convective 
envelope being in agreement with the result of helioseismic inversions 
($R_{CZ}=0.713R_{\odot}$), the initial metal abundance has to be fixed to 
$Z_0=0.0208$ which results in a higher $Z/X$ (0.0255) compared to the recent 
observations (0.0181) \citep{asp}.

\begin{deluxetable*}{cccccccccc}
\tabletypesize{\scriptsize}
\tablecaption{Parameters of the convection model and properties of solar models}
\tablewidth{0pt}
\tablehead{ \colhead{Model}        & \colhead{$c_{\mu}$} & 
            \colhead{${c'}_{\mu}$} & \colhead{$c_T$}     & 
            \colhead{$\alpha$}     & \colhead{$Y_0$}     & 
            \colhead{$Y_S$}        & \colhead{$Z_0$}     & 
            \colhead{$Z/X$}        & \colhead{$R_{CZ}(R_{\odot})$}}
\startdata
MLT &     &       &     & 1.8750 & 0.27785 & 0.2474 & 0.0208 & 0.0255 & 0.7136\\
STC & 0.1 & 0.008 & 0.5 & 1.5536 & 0.27785 & 0.2474 & 0.0208 & 0.0255 & 0.7138
\enddata
\end{deluxetable*}

\begin{figure}
\epsscale{1.2}
\plotone{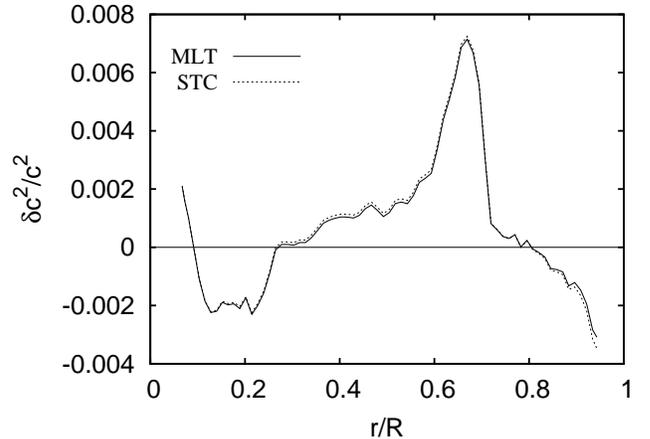}
\caption{The sound speed difference between the helioseismic inversion and solar models based on different convection models. The solid line (MLT) represents the result of the standard mixing-length theory, and the dashed line (STC) is the result of the steady state solution of the $k$-$\omega$ model in the fully local equilibrium. }
\end{figure}
\begin{figure}
\epsscale{1.2}
\plotone{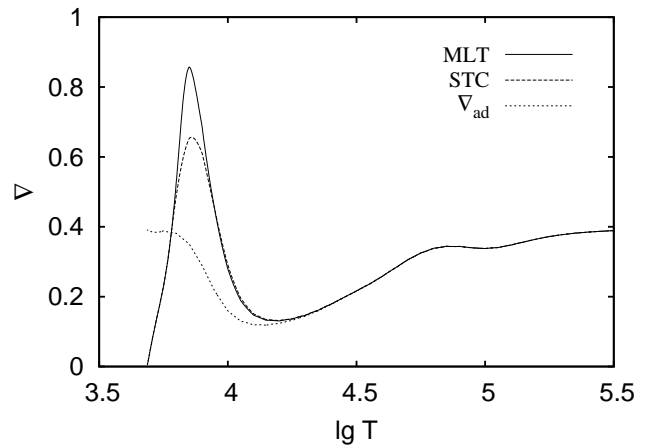}
\caption{The temperature gradient in the upper envelope of solar models based on different convection models. The solid line (MLT) represents the result of the standard mixing-length theory, and the dashed line (STC) is the result of the steady state solution of the $k$-$\omega$ model in the fully local equilibrium. The adiabatic temperature gradient is given as the dotted line for reference. }
\end{figure}

The sound speed from the helioseismic inversion \citep{bas} is compared with 
those of our solar models with the mixing-length theory (MLT) and with the 
turbulent convection model of Eq. (94) (STC) in Fig. 2. It can be seen that the 
two solar models have almost identical sound speed profiles, except for in a 
region about 5\% below the solar photosphere. In order to show the difference 
of the two solar models just below the solar photosphere, the profiles of the 
temperature gradient $\nabla$ of MLT and of STC are compared in Fig. 3, along 
with the adiabatic temperature gradient $\nabla_{ad}$ for reference. It can be 
seen that the turbulent convection model significantly reduces the temperature 
gradient just below the photosphere, which is a direct result of lower 
decreasing rate of the convective heat transfer efficiency of the turbulent 
convection model \citep{tr10}. 

The typical velocity of turbulent motions in the convection zone is compared 
in Fig. 4 for the two solar models with MLT and STC. It can be noticed that 
the two turbulence models predict similar maximum velocities of turbulence 
(about $3\times{10^5}\,{\rm cm\,s^{-1}}$) appeared just below the solar 
photosphere. But in most part of the convection zone, the STC predicts a 
typical velocity of turbulence (about $10^3\,{\rm cm\,s^{-1}}$) one order of 
magnitude lower than what the MLT does (about $10^4\,{\rm cm\,s^{-1}}$).

\begin{figure}
\epsscale{1.2}
\plotone{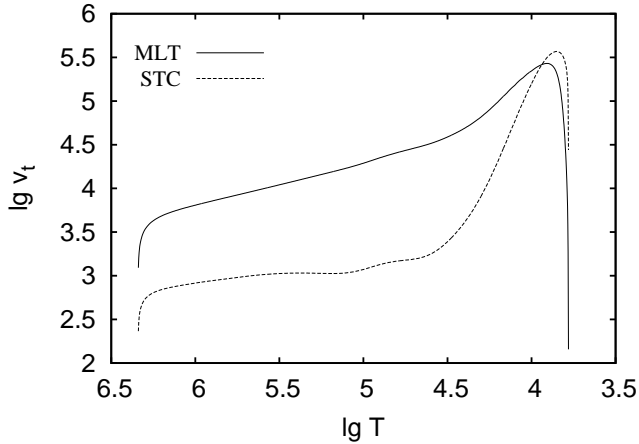}
\caption{The distribution of the typical velocity of turbulence based on different convection models in the solar convection zone. The solid line (MLT) represents the result of the standard mixing-length theory, and the dashed line (STC) is the result of the steady state solution of the $k$-$\omega$ model in the fully local equilibrium. }
\end{figure}

Secondly, we apply the full $k$-$\omega$ model of Eqs. (104) and (109) to the
whole solar interior including the convection zone and overshooting regions 
located below and above it, in order to study the transport effect of 
turbulence. It can be seen in Fig. 5 that the transport effect of the 
gradient diffusion turbulence model (GDT) is significant near the boundaries 
of the convection zone and in the overshooting regions. The typical velocity 
of turbulence roughly shows a linearly decaying law below the base of the 
solar convection zone, and according to results of different combinations of 
parameters values of the stellar turbulence model, the larger the parameter 
$c'_{\mu }$ is, the steeper the decaying law will be, which requires a smaller 
$\alpha$ to calibrate the model to the solar values. However, the turbulent 
velocity extends more or less constantly into the solar photosphere and lower 
atmosphere until they drop abruptly to the outer boundary conditions we have 
prescribed in the upper atmosphere. It is interesting to note that the typical 
velocity of turbulence is restricted below $1\,{\rm km\,s^{-1}}$ in the whole 
subphotospheric region due to the transport effect of turbulence, ensuring the 
thermally turbulent convection appearing there a subsonic motion as assumed in 
the stellar turbulence model. On the other hand, the typical length of 
turbulent motion remains almost a constant in the overshooting region below the 
base of the solar convection zone as shown in Fig. 6. Contrary to the local 
pressure scale height that decreases monotonically in the stellar envelope, the 
typical length of turbulence increases exponentially in the upper convection 
zone and overshooting region. This is a direct result of radiative dissipation
that effectively increases the average size of convective rolling cells when 
the density drops rapidly in the solar convective envelope.

\begin{figure}
\epsscale{1.2}
\plotone{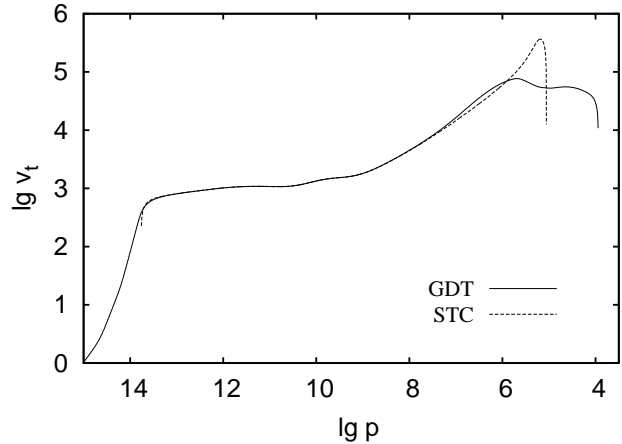}
\caption{The comparison of the typical velocity of turbulence based on the steady state solution in the fully local equilibrium (STC) and on the full $k$-$\omega$ model including the transport effect with the gradient diffusion turbulence model (GDT). }
\end{figure}
\begin{figure}
\epsscale{1.2}
\plotone{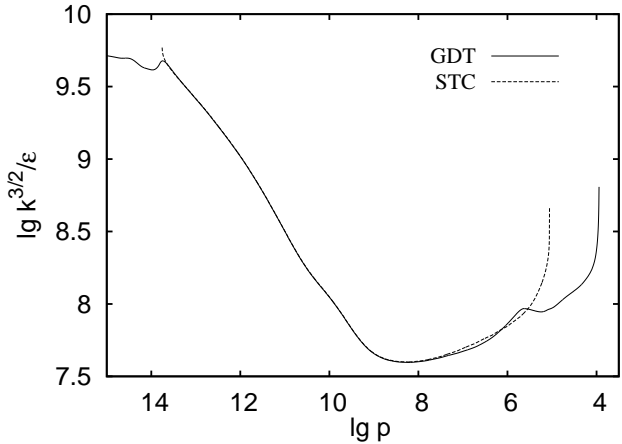}
\caption{The comparison of the typical length of turbulence based on the steady state solution in the fully local equilibrium (STC) and on the full $k$-$\omega$ model including the transport effect with the gradient diffusion turbulence model (GDT). }
\end{figure}

\subsection{Comparisons of stellar evolutionary models}

We calculate a series of stellar evolution models of $0.6M_{\odot}$, 
$1.0M_{\odot}$, $1.5M_{\odot}$, $3.0M_{\odot}$, and $7.0M_{\odot}$, with the 
initial chemical abundance of $Y_0=0.27785$ and $Z_0=0.02$ and parameters of
turbulence models given in Table 2. Element diffusion is not considered here 
for simplicity. The HR diagram is shown in Fig. 7 for the stellar models with 
different masses. It can be seen that the evolution tracks of stellar models 
based on the mixing-length theory (MLT) and on the model of stellar turbulent 
convection (STC) are almost identical, except that the STC results in a series 
of red giant branches with a little bit lower effective temperatures.

\begin{figure*}
\epsscale{0.8}
\plotone{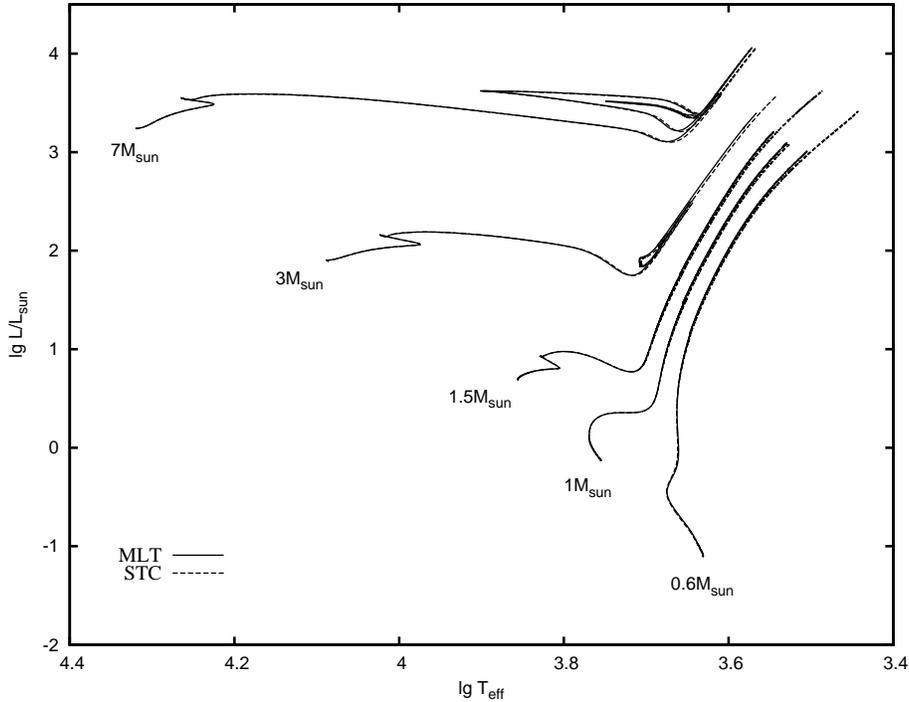}
\caption{The HR diagram for stellar models of different masses based on different convection models. The solid line (MLT) represents the result of the standard mixing-length theory, and the dashed line (STC) is the result of the steady state solution of the $k$-$\omega$ model in the fully local equilibrium. }
\end{figure*}
\begin{figure}
\epsscale{1.2}
\plotone{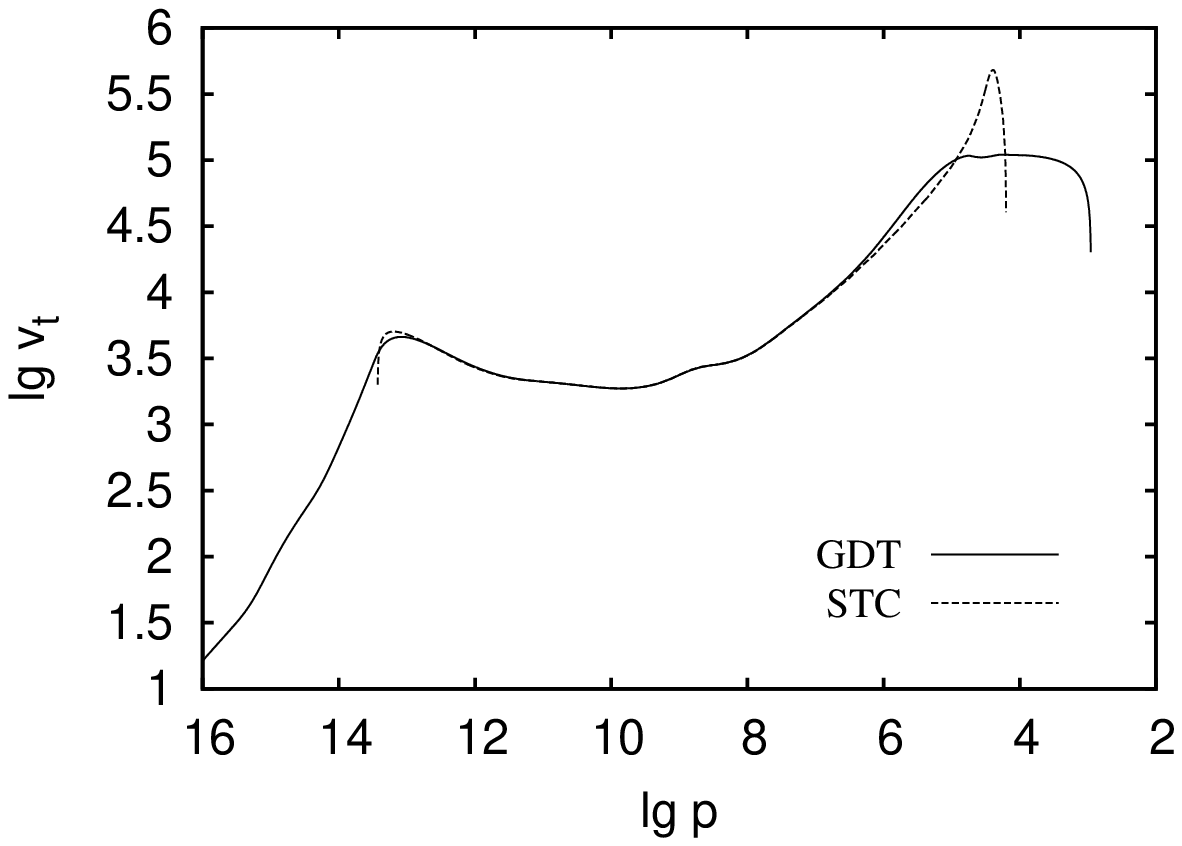}
\caption{The comparison of the typical velocity of turbulence for an $1.0M_{\odot}$ RGB model. The solid line represents the result of the full $k$-$\omega$ model including the transport effect with the gradient diffusion turbulence model (GDT), and the dashed line is the result of the steady state solution in the fully local equilibrium (STC). }
\end{figure}
\begin{figure}
\epsscale{1.2}
\plotone{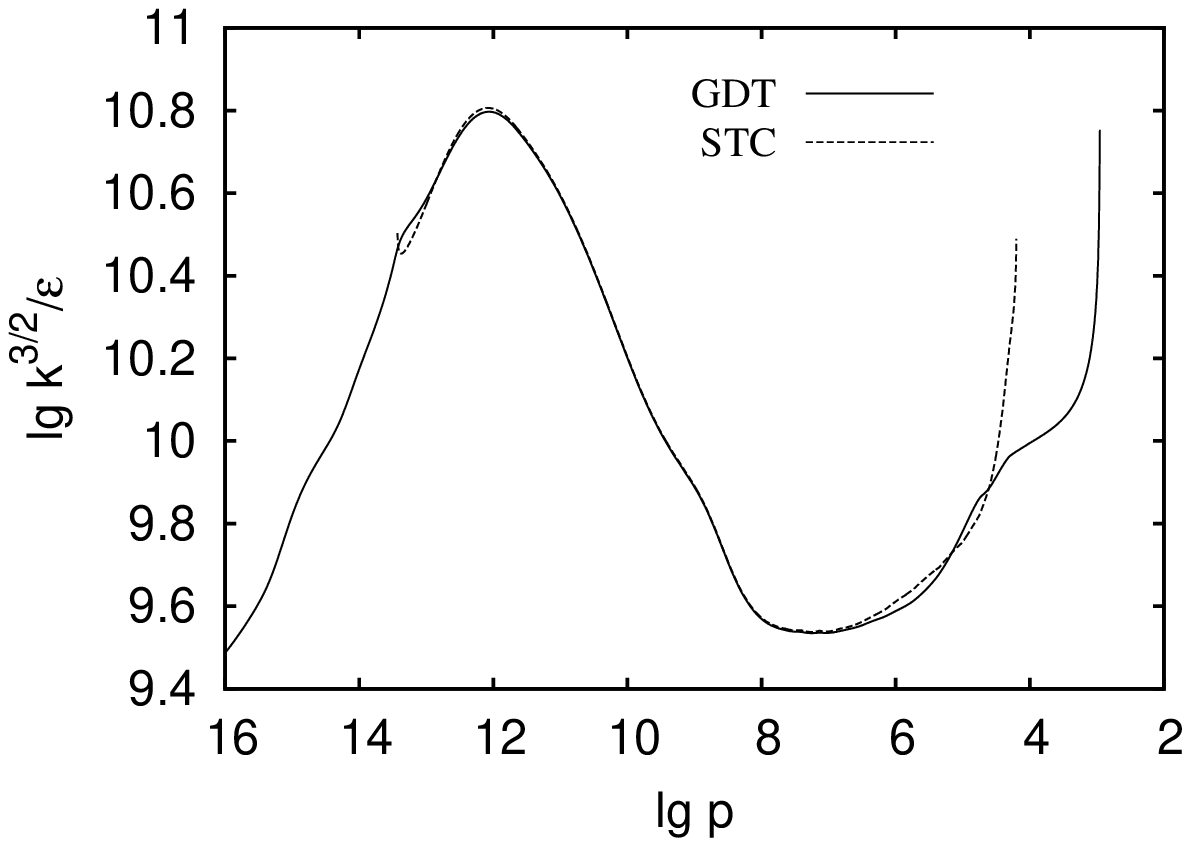}
\caption{The comparison of the typical length of turbulence for an $1.0M_{\odot}$ RGB model. Others are the same as in Fig. 8. }
\end{figure}

Profiles of the typical velocity of turbulence are compared for $1.0M_{\odot}$ 
models based on STC and GDT at the RGB bump in Fig. 8, within which the 
envelope convection mostly penetrates inwardly into the stellar interior. It 
can be seen that the transport effect of turbulence is only significant below 
the base of the convective envelope, with an almost linearly decaying law 
similar to the solar case. On the other side, the turbulent velocity is 
effectively reduced in the upper convective envelope and diffused as nearly 
a constant value of $1\,{\rm km\,s^{-1}}$ into the stellar atmosphere. It is 
interesting to note in Fig. 9, however, that the typical length of turbulence 
varies only in one order of magnitude around $10^5\,$km in the whole convective 
envelope, decreasing linearly below the base of the convective envelope and 
increasing rapidly in the stellar atmosphere.

For a model of $7.0M_{\odot}$ in the main sequence, the transport effect of
turbulence results in significant effect on the turbulent velocity in the 
convective core, the maximum velocity of GDT being only 80\% as seen in 
Fig. 10 as that of STC. Outside the convective core, the turbulent velocity 
can be diffused into an overshooting region of about $0.5H_p$ wide. However, 
the turbulent velocity is decaying more and more rapidly in the overshooting 
region, resulting in a considerably shorter distance of fully mixing for 
elements outside the convective core. It can be seen in Fig. 11, however, 
that the typical length of turbulence remains almost as a constant in the 
convective core and in the overshooting region, and its value is comparable 
to the size of the convective core. The typical length of turbulence sharply 
increases toward the stellar center due to the central boundary condition, 
which assumes the local equilibrium of turbulence at the center of a star.

\begin{figure}
\epsscale{1.2}
\plotone{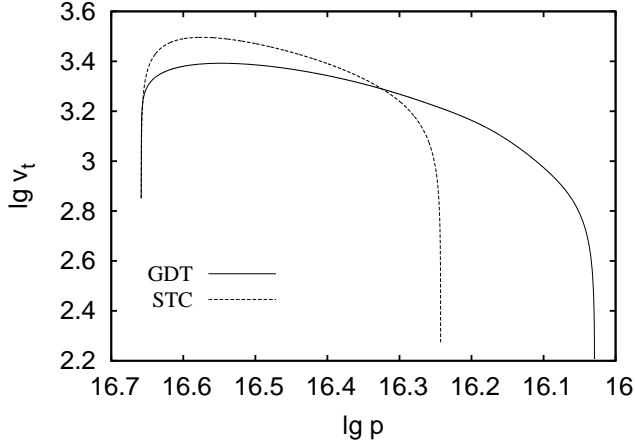}
\caption{The comparison of the typical velocity of turbulence for a $7.0M_{\odot}$ main sequence model. The solid line represents the result of the full $k$-$\omega$ model including the transport effect with the gradient diffusion turbulence model (GDT), and the dashed line is the result of the steady state solution in the fully local equilibrium (STC). }
\end{figure}
\begin{figure}
\epsscale{1.2}
\plotone{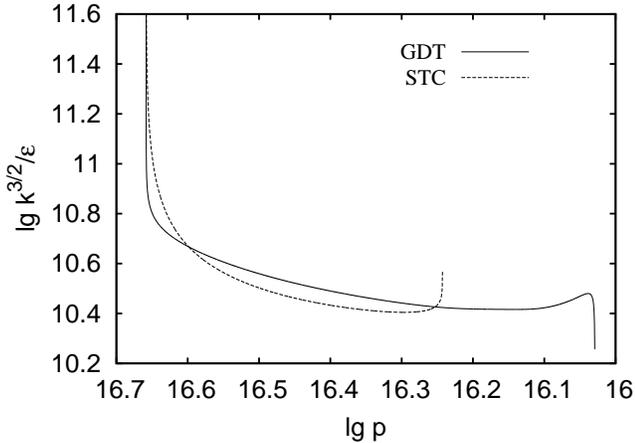}
\caption{The comparison of the typical length of turbulence for a 
$7.0M_{\odot}$ main sequence model. Others are the same as in Fig. 10. }
\end{figure}

Finally we apply the turbulence models of STC and GDT to a $3.0M_{\odot}$ 
model in the post-main-sequence evolutionary stage. According to the MLT,
convection is underdeveloped in the envelope of such a star and turbulent 
motions only appear in two thin shells and have almost no effect on the 
structure of the stellar envelope. As seen in Fig. 12, however, the transport 
effect dominates the turbulent motion, and the turbulent velocity is diffused 
in the whole upper envelope upward to the stellar atmosphere and downward 
several local pressure scale heights below the lowest convective boundary. 
The resulted maximum velocity of turbulence is substantially reduced from 
STC's $16\,{\rm km\,s^{-1}}$ to GDT's $4\,{\rm km\,s^{-1}}$. On the other hand, 
the typical length of turbulence remains almost a constant as seen in Fig. 13, 
and its value is significantly increased to be comparable to the width of the 
convective region being formed thereabout by the turbulent diffusion.

\begin{figure}
\epsscale{1.2}
\plotone{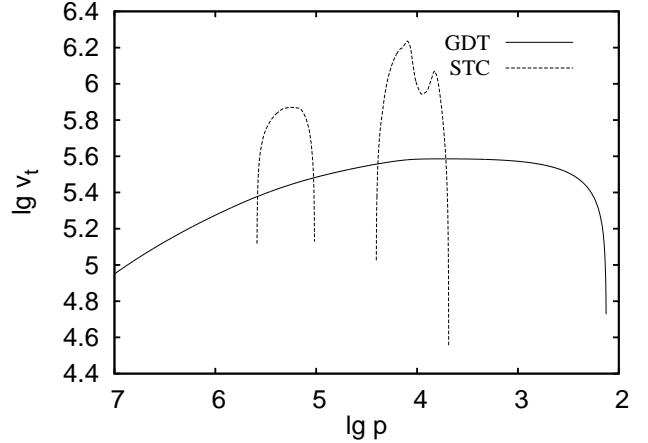}
\caption{The comparison of the typical velocity of turbulence for a $3.0M_{\odot}$ sub-giant model. The solid line represents the result of the full $k$-$\omega$ model including the transport effect with the gradient diffusion turbulence model (GDT), and the dashed line is the result of the steady state solution in the fully local equilibrium (STC). }
\end{figure}
\begin{figure}
\epsscale{1.2}
\plotone{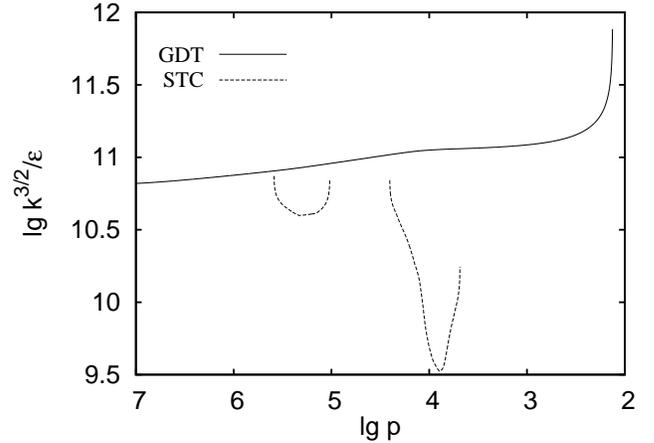}
\caption{The comparison of the typical length of turbulence for a 
$3.0M_{\odot}$ sub-giant model. Others are the same as in Fig. 12. }
\end{figure}

\section{ Conclusions and Discussions }

Thermal convection is characterized in stars by convective rolling cells of 
different scales and a fully developed turbulence. Such semi-regular and large 
scale structures play a critical role in determining the macro properties of 
the convective motions, by both acting as thermal engines to transform heat 
into the kinetic energy to maintain their cellular configurations and producing 
the velocity shear within the convective rolling cells to generate turbulence. 
Due to great difficulties encountered both physically and numerically in 
dealing with such complex flows, turbulent convection models, which are based 
on fully hydrodynamic moment equations and include many physical properties of 
turbulence, have not jet taken the contributions from the convective rolling 
cells into account. 

In order to investigate the qualitative properties of the convective rolling 
cells and to include their contributions to the development of turbulence, we 
simplify their configurations by eddies rotating circle by circle with 
differential angular velocities, and approximately describe their motions by 
equations of mass and energy conservation. With the aid of advanced turbulence 
models, this simple approach enable us to estimate the average size of the 
convective rolling cells by solving for the temperature difference over the 
stationary temperature background, and to approximate the generation rate of 
turbulence due to the average shear of velocity in the convective rolling 
cells. On the other hand, the restriction from the heat conversion efficiency 
of such convective rolling cells working as individual thermal engines leads 
to a direct limit on the turbulent kinetic energy enclosed in them and an 
indirect evaluation of the typical length scale of turbulence in the fully 
local equilibrium. Based on these considerations and model assumptions thereof, 
we obtain the static solution of the standard $k$-$\varepsilon$ model by 
properly introducing an important model parameter $c_{\varepsilon 3}$, and 
then develop a $k$-$\omega$ model including the turbulent transport effect 
to describe stellar turbulent convection. Results of preliminary applications 
to the sun and other stars with different masses and in different evolutionary 
stages are in good agreement with those of the standard mixing-length theory 
and numerical simulations.

Being as an initial attempt to include the contributions from the convective 
rolling cells in the turbulent convection models, our approach have a lot of 
limitations and incongruities. The most significant among them lies in the 
simplistic assumption that the convective rolling cells are of an eddy-like 
configuration, which is proved to be inadequate for the surface layer 
convection in the solar convective envelope where the scale height of the 
density stratification is much smaller than the average size of the convective 
rolling cells. The huge variation of density in the convective rolling cells 
results in great changes on the topology of the flows, i.e. a distinct up/down 
asymmetry characterized by hot and slow up-flows separated by cold and fast 
down-drafts \citep{sn89}. On the other hand, the properties of the convective 
flow is determined by the physical conditions along the whole streamline that 
the fluids go through during a complete cycle of the convective motion. Such a 
nonlocal nature of the phenomenon, along with the huge range of variations of 
the density, significantly limits the validity of our approach by use of the 
local physical conditions to estimate the size of the convective rolling cells. 
Besides, the actual size and shape of the convective rolling cells may also be 
restricted by geometric conditions, for examples, in the stellar convective 
core or in thin convective shells of the stellar envelope with a thickness 
smaller than the local size of the convective rolling cells.

\acknowledgments 

This work is sponsored by the NSFC of China through project number 10673030 
and 10973035. Fruitful discussions with Q.S. Zhang are highly appreciated.

\clearpage

\end{document}